\documentclass{amsart}
\usepackage{amsfonts}
\usepackage{youngtab}
\usepackage[pagebackref]{hyperref}

\theoremstyle{plain}

\numberwithin{equation}{section}
\newcommand{\qbin}[2]{\genfrac{[}{]}{0pt}{}{#1}{#2}}
\newcommand{\qfloor}[2]{\genfrac{\lfloor}{\rfloor}{0pt}{}{#1}{#2}}
\input{tcilatex}

\begin{document}
\title[Chern-Simons theory and Stieltjes-Wigert polynomials]{Chern-Simons
matrix models and Stieltjes-Wigert polynomials}
\author{Yacine Dolivet}
\address{Laboratoire de Physique Th\'{e}orique de l'\'{E}cole Normale Sup%
\'{e}rieure. 24 rue L'homond 75231, Paris Cedex 05, France.}
\email{dolivet@lpt.ens.fr}
\author{Miguel Tierz}
\address{Institut d'Estudis Espacials de Catalunya (IEEC/CSIC). Campus UAB,
Facultat de Ci\'{e}ncies, Torre C5-Parell-2a planta. E-08193 Bellaterra
(Barcelona) Spain.}
\email{tierz@ieec.fcr.es}

\begin{abstract}
Employing the random matrix formulation of Chern-Simons theory on Seifert
manifolds, we show how the Stieltjes-Wigert orthogonal polynomials are
useful in exact computations in Chern-Simons matrix models. We construct a
biorthogonal extension of the Stieltjes-Wigert polynomials, not available in
the literature, necessary to study Chern-Simons matrix models when the
geometry is a lens space. We also discuss several other results based on the
properties of the polynomials: the equivalence between the Stieltjes-Wigert
matrix model and the discrete model that appears in $q$-2D\ Yang-Mills and
the relationship with Rogers-Szeg\"{o} polynomials and the corresponding
equivalence with an unitary matrix model. Finally, we also give a detailed
proof of a result that relates quantum dimensions with averages of Schur
polynomials in the Stieltjes-Wigert ensemble.
\end{abstract}

\maketitle
%\tableofcontents

%\begin{flushright}
%LPTENS-06/44\newline
%\end{flushright}

\section{Introduction}

In the late eighties \cite{cs}, Witten considered a topological gauge theory
for a connection on an arbitrary three-manifold $M,$ based on the
Chern-Simons action: 
\begin{equation}
S_{\mathrm{CS}}(A)={\frac{k}{4\pi}}\int_{M}\mathrm{Tr}(A\wedge dA+{\frac{2}{3%
}}A\wedge A\wedge A),  \label{cs}
\end{equation}
with $k$ an integer number. One of the most important aspects of
Chern-Simons theory is that it provides a physical approach to three
dimensional topology. In particular, it gives three-manifold invariants and
knot invariants. For example, the partition function,

\begin{equation}
Z_{k}(M)=\int\mathcal{D}A\mathrm{e}^{iS_{\mathrm{CS}}(A)},  \label{wrt}
\end{equation}
delivers a topological invariant of $M$, the so-called
Reshetikhin-Turaev-Witten invariant. Recent reviews are \cite%
{Marino:2004eq,Marino:2004uf}.

As reviewed in \cite{Marino:2004uf}, a great deal of interest has focused on
the fact that Chern-Simons theory provides large $N$ duals of topological
strings. This connection between Chern-Simons theory and topological strings
was already pointed out by Witten \cite{wittenopen} (see also \cite{Periwal}%
), and then extended in \cite{Gopakumar:1998ki}.

Recent progress in Chern-Simons theory includes a description of
Chern-Simons theory on certain geometries in terms of models of random
matrices. Consider the partition function of Chern-Simons theory on a
Seifert space $M=X({\frac{p_{1}}{q_{1}}},\ldots ,{\frac{p_{n}}{q_{n}}})$.
This is obtained by doing surgery on a link in $S^{3}$ with $n+1$
components, out of which $n$ are parallel, unlinked unknots, and one has
link number $1$ with each of the $n$ unknots. The surgery data are $%
p_{j}/q_{j}$ for the unlinked unknots, $j=1,\ldots ,n$, and 0 for the last
component. The partition function is\footnote{%
See the Appendix C for details on the notation} \cite{Marino:2002fk}: 
\begin{align}
\mathbf{Z}_{\mathrm{CS}}(M)& ={\frac{(-1)^{|\Delta _{+}|}}{|\mathcal{W}%
|\,(2\pi i)^{r}}}\Biggl({\frac{\mathrm{Vol}\,\Lambda _{\mathrm{w}}}{\mathrm{%
Vol}\,\Lambda _{\mathrm{r}}}}\Biggr){\frac{[\mathrm{sign}(P)]^{|\Delta _{+}|}%
}{|P|^{r/2}}}\mathrm{e}^{{\frac{\pi id}{4}}\mathrm{sign}(H/P)-{\frac{\pi idy%
}{12l}}\phi }  \label{20} \\
& \times \sum_{t\in \Lambda _{\mathrm{r}}/H\Lambda _{\mathrm{r}}}\int d\beta
\,\mathrm{e}^{-{\beta ^{2}/2g_{s}}-lt\cdot \beta }{\frac{\prod_{i=1}^{n}%
\prod_{\alpha >0}2\sinh {\frac{\beta \cdot \alpha }{2p_{i}}}}{\prod_{\alpha
>0}\left( 2\sinh {\frac{\beta \cdot \alpha }{2}}\right) ^{n-2}}}~.  \notag
\end{align}%
This expression gives the contribution of the reducible flat connections to
the partition functions. Recall that for both $S^{3}$ and lens spaces this
amounts to the exact partition function. The case $n=0$ corresponds to the
three-sphere $S^{3}$ that leads to $\left( \ref{sinh}\right) .$ Thus, for
the case of $U(N)$, and focusing on a particular sector of flat connections,
we get the following matrix model: 
\begin{equation}
Z_{\mathrm{CS}}(M)=\prod_{i=1}^{N}\int_{-\infty }^{\infty }{\mbox{d}}y_{i}\,%
\mathrm{e}^{-y_{i}^{2}/2g_{s}-lt_{i}y_{i}}{\frac{\prod_{j=1}^{n}\prod_{k<l}2%
\sinh {\frac{y_{k}-y_{l}}{2p_{j}}}}{\prod_{k<l}\left( 2\sinh {\frac{%
y_{k}-y_{l}}{2}}\right) ^{n-2}}}~.  \label{genUn}
\end{equation}%
Of course, the simplest case is that of $S^{3}$ with gauge group $U(N),$
which is given by the partition function of the following random matrix
model: 
\begin{equation}
Z=\frac{\mathrm{e}^{-\frac{g_{s}}{12}N\left( N^{2}-1\right) }}{N!}\int
\prod_{i=1}^{N}\mathrm{e}^{-u_{i}^{2}/2g_{s}}\prod_{i<j}\left( 2\sinh \frac{%
u_{i}-u_{j}}{2}\right) ^{2}\frac{du_{i}}{2\pi }.  \label{sinh}
\end{equation}%
From the point of view of topological strings, this describes open
topological $A$ strings on $T^{\ast }\emph{S}^{3}$ with $N$ branes wrapping $%
S^{3}$ \cite{Marino:2002fk}. This latter case, as shown in \cite{Tierz}, can
be studied with usual techniques of random matrix theory. More precisely,
the Stieltjes-Wigert polynomials, a member of the $q$-deformed orthogonal
polynomials family \cite{Koe}, allows to compute, in exact fashion,
quantities associated to the matrix model. In the computation, the $q$%
-parameter of the polynomials turns out to be naturally identified with the $%
q$-parameter of the quantum group invariants associated to the Chern-Simons
theory. This is so because the previous model can be easily mapped into:

\begin{equation}
Z=\int [dM]\mathrm{e}^{-{\frac{1}{2g_{s}}}{\mathrm{Tr}}(\log M)^{2}}~,
\label{SW}
\end{equation}%
named Stieltjes-Wigert ensembles, after the associated orthogonal
polynomials.

Chern-Simons matrix models have been further considered in \cite{TdH1} and 
\cite{Aga}-\cite{deHaro:2004id} and also play a central role in $q$-2D
Yang-Mills theory \cite{Aganagic:2004js}-\cite{Caporaso:2005np}. Most of
these works focus on the relevance to topological strings. In \cite%
{Tierz,TdH1}, the emphasis is on exact solutions and on the special features
of the matrix models. The works of Caporaso et al. \cite%
{Caporaso:2005fp,Caporaso:2005ta,Caporaso:2005np} also make an extensive use
of the properties of the Stieltjes-Wigert orthogonal polynomials. We shall
be focussing here on aspects of the Chern-Simons matrix models that have to
do with the associated system of orthogonal polynomials.

This paper is organized as follows. In the next section we shall construct a
biorthogonal extension of the Stieltjes-Wigert polynomials, in order to
study the matrix model $\left( \ref{genUn}\right) $ when $n=1$ and $n=2$.
These polynomials have not been discussed in the (vast) orthogonal
polynomials literature, so most of our effort is on their derivation and to
establish some of its fundamental properties. They are necessary if one
wants to obtain full analytic results when the geometry is something more
complicated than $S^{3}.$ Note that matrix models in the lens space case
have already been studied (with loop equations) \cite{HOY}, but if one
desires an all order result as in \cite{Tierz}, the knowledge of orthogonal
polynomials is then necessary. After the construction of the biorthogonal
Stieltjes-Wigert polynomials in Section 2, we discuss some of their
mathematical properties in Section 3. In the last Section, we discuss
several aspects of the Chern-Simons matrix models by focussing exclusively
on properties of the (ordinary) Stieltjes-Wigert polynomials. In particular,
we clearly establish the relationship with the discrete matrix model that
also appears in $q$-2D Yang-Mills theory, and also employ the intimate
relationship between Stieltjes-Wigert and Rogers-Szeg\"{o} polynomials to
find the exact relation between the Stieltjes-Wigert matrix model and
Okuda's unitary matrix model \cite{Okuda:2004mb}. Finally, we give a
detailed proof, employing a mixture of combinatorial and orthogonal
polynomials results, of the equality between quantum dimensions and averages
of Schur polynomials in the Stieltjes-Wigert ensemble \cite{Marino:2004eq}.
We conclude with a summary and with some avenues for further research,
presented in the Conclusions and Outlook.

\section{Biorthogonal Stieltjes-Wigert}

Let us consider the generic expression $\left( \ref{genUn}\right) $ in the $%
n=1$ and $n=2$ cases, that correspond to the case of lens spaces. We are
lead to a biorthogonal extension of the $S^{3}$ model:%
\begin{equation}
Z=\int \prod_{i=1}^{N}\mathrm{e}^{-u_{i}^{2}/2g_{s}}\prod_{i<j}\left( 2\sinh 
\frac{u_{i}-u_{j}}{2P}\right) \left( 2\sinh \frac{u_{i}-u_{j}}{2Q}\right) 
\frac{du_{i}}{2\pi }.  \label{lens}
\end{equation}%
Recall that a biorthogonal ensemble of random matrices has the probability
density \cite{Bor}:

\begin{equation}
P(x_{1},...,x_{N})=C_{N}\dprod\limits_{i=1}^{N}\omega \left( x_{i}\right)
\dprod\limits_{i<j}\left( x_{i}-x_{j}\right) \left(
x_{i}^{k}-x_{j}^{k}\right) ,  \label{biorthogonal}
\end{equation}%
where $k$ is a fixed real number. In total analogy with the usual Hermitian
case ($k=1$ ) one can study $\left( \ref{biorthogonal}\right) $ by
considering a pair of biorthogonal polynomials:%
\begin{equation}
\int \omega \left( x\right) Y_{n}\left( x,k\right) Z_{m}\left( x,k\right)
dx=h_{n,k}\delta _{n,m},
\end{equation}%
with:%
\begin{eqnarray}
\int Y_{n}\left( x,k\right) x^{kj}\omega \left( x\right) dx &=&\alpha
_{n}^{\left( k\right) }\delta _{n,j}, \\
\int Z_{n}\left( x,k\right) x^{j}\omega \left( x\right) dx &=&\beta
_{n}^{\left( k\right) }\delta _{n,j}.  \notag
\end{eqnarray}%
We warn the reader that the term \textit{biorthogonal} is employed in
different contexts in the literature. The classical cases (Hermite, Laguerre
and Jacobi) were worked out in \cite{Bor}. Note that $\left( \ref%
{biorthogonal}\right) $ is exactly the type of ensemble that $\left( \ref%
{lens}\right) $ leads us to consider since:

\begin{eqnarray}
Z^{P,Q} &=&\int \prod_{i}{\frac{\mbox{d}u_{i}}{2\pi }}\mathrm{e}%
^{-u_{i}^{2}/2g_{s}}\prod_{i<j}(2\sinh ({\frac{u_{i}-u_{j}}{2P}}))(2\sinh ({%
\frac{u_{i}-u_{j}}{2Q}})  \notag \\
&=&q^{-{\frac{N\alpha ^{2}}{2}}}\int \prod_{i}{\frac{\mbox{d}y_{i}}{2\pi }}%
\mathrm{e}^{-{\kappa }^{2}\log
^{2}y_{i}}\prod_{i<j}(y_{i}^{1/P}-y_{j}^{1/P})(y_{i}^{1/Q}-y_{j}^{1/Q}),
\end{eqnarray}%
with $u_{i}=\log \mathrm{e}^{{\frac{\alpha }{2{\kappa }^{2}}}}{y_{i}},$ $%
\kappa ^{2}=1/2g_{s}$ and $\alpha =-1-{\frac{\beta (N-1)}{2}},$ $\beta ={%
\frac{1}{P}}+{\frac{1}{Q}.}$ Finally, with $y_{i}=\mathrm{e}^{{\frac{P-1}{2{%
\kappa }^{2}P}}}x_{i}^{P}$ and some rewriting:

\begin{equation}
Z^{P,Q}=P^{N}\mathrm{e}^{-{\frac{N}{4{\kappa }^{2}}}(\frac{1}{P}+{\frac{%
\beta (N-1)}{2}})^{2}}\int \prod_{i}{\frac{\mbox{d}w_{i}}{2\pi }}\mathrm{e}%
^{-\kappa ^{2}P^{2}\log
^{2}x_{i}}\prod_{i<j}(x_{i}-x_{j})(x_{i}^{P/Q}-x_{j}^{P/Q}),  \label{eq:ZPQ}
\end{equation}%
which is of the form $\left( \ref{biorthogonal}\right) $ with the log-normal
(Stieltjes-Wigert) weight function: $\omega \left( x\right) =\mathrm{e}%
^{-\kappa ^{2}P^{2}\log ^{2}x_{i}}$, the $q$-parameter is then $q=\mathrm{e}%
^{-\frac{1}{2\kappa ^{2}P^{2}}}=\mathrm{e}^{-\frac{g_{s}}{P^{2}}}.$
Therefore, if we want to go beyond the $S^{3}$ case one has to construct the
biorthogonal Stieltjes-Wigert polynomials, not available in the literature.
Thus, this is our main task in what follows. The method we have chosen is
based on a simple but fundamental result by Askey, that relates the $q$%
-Laguerre orthogonal polynomials and the Stieltjes-Wigert polynomials \cite%
{Askey}:

\begin{equation}
\lim_{\alpha \rightarrow \infty }L_{n}^{\alpha }\left( q^{-\alpha
}x;q\right) =S_{n}(x;q),  \label{Askey}
\end{equation}%
and then we take into account the biorthogonal construction of the $q$%
-Laguerre polynomials, carried out by Al-Salam and Verma in the early
eighties \footnote{%
The resulting polynomials were named $q$-Konhauser as they could also be
interpreted as a $q$-deformed version of the biorthogonal Laguerre
polynomials, worked out by Konhauser.}. The Stieltjes-Wigert polynomials are 
\cite{Szego}\footnote{%
In \cite{Askey} they appear, in\ Eq. (2.5), slightly reformulated.}%
\begin{equation}
S_{n}(x|q)\equiv {\frac{1}{(q;q)_{n}}}\sum_{r=0}^{n}\QATOPD[ ] {n}{r}%
_{q}(-1)^{r}q^{r^{2}}x^{r}.  \label{eq:SW}
\end{equation}

The limit $\left( \ref{Askey}\right) $ will provide us with a biorthogonal
extension of the SW polynomials starting with the $q$-Konhauser polynomials.
Therefore, following \cite{AlSalam}, let us write:

\begin{equation}
Z_{n}^{(\alpha)}(x,k|q)\equiv\frac{\lbrack q^{1+\alpha}]_{nk}}{%
(q^{k};q^{k})_{n}}\sum_{j=0}^{n}\frac{(q^{-nk};q^{k})_{j}q^{\frac{1}{2}%
kj(kj-1)+kj(n+\alpha+1)}}{(q^{k};q^{k})_{j}[q^{1+\alpha}]_{kj}}x^{kj},
\end{equation}
and

\begin{equation}
Y_{n}^{(\alpha)}(x,k|q)\equiv\frac{1}{[q]_{n}}\sum_{r=0}^{n}\frac {x^{r}q^{%
\frac{1}{2}r(r-1)}}{[q]_{r}}b_{r}^{\alpha},
\end{equation}
with

\begin{equation}
b_{r}^{\alpha}\equiv\sum_{s=0}^{r}\frac{[q^{-r}]_{s}}{[q]_{s}}%
q^{s}(q^{1+\alpha+s};q^{k})_{n}.
\end{equation}
These polynomials satisfy: 
\begin{equation}
<Z_{n}^{(\alpha)}(x,k|q),Y_{m}^{(\alpha)}(x,k|q)>=k_{n}^{(\alpha)}\delta
_{n,m}\qquad\text{with\qquad}k_{n}^{(\alpha)}=\frac{[q^{1+%
\alpha}]_{nk}q^{-nk}}{[q]_{n}},
\end{equation}
with respect to the normalized $q$-Laguerre measure. We have to study:

\begin{align}
Z_{n}(x,k|q) & \equiv\lim_{\alpha\rightarrow\infty}Z_{n}^{(\alpha
)}(q^{-\alpha}x,k|q), \\
Y_{n}(x,k|q) & \equiv\lim_{\alpha\rightarrow\infty}Y_{n}^{(\alpha
)}(q^{-\alpha}x,k|q).  \notag
\end{align}
In the first case, one readily finds:

\begin{equation}
Z_{n}(x,k|q)=\frac{1}{(q^{k};q^{k})_{n}}\sum_{j=0}^{n}\frac{%
(q^{-nk};q^{k})_{j}q^{\frac{1}{2}kj(kj-1)+kj(n+1)}}{(q^{k};q^{k})_{j}}x^{kj},
\end{equation}%
which can be conveniently reexpressed: %thanks to Eq.\ref{eq:invert}

\begin{equation}
\boxed{ Z_n(x,k|q)
=\frac{1}{(q^k;q^k)_n}\sum_{r=0}^{n}\qbin{n}{r}_{q^k}(-1)^rq^{{1\over
2}r^2k(k+1)} x^{kr}.}  \label{eq:Z}
\end{equation}%
Regarding $Y_{n}(x,k|q)$ we have to find $b_{r}\equiv \lim_{\alpha
\rightarrow \infty }q^{-\alpha r}b_{r}^{\alpha }.$ Employing $q$-Taylor \cite%
{AlSalam} one can write:

\begin{equation}
(q^{1+\alpha}x;q^{k})_{n}=\sum_{r=0}^{n}\frac{x^{r}[1/x]_{r}}{[q]_{r}}%
\sum_{s=0}^{r}\frac{[q^{-r}]_{s}}{[q]_{s}}q^{s}(q^{1+\alpha+s};q^{k})_{n},
\end{equation}
therefore:

\begin{equation}
(qx;q^{k})_{n}=\sum_{r=0}^{n}\frac{x^{r}[q^{\alpha }/x]_{r}}{[q]_{r}}%
q^{-\alpha r}b_{r}^{\alpha }.
\end{equation}%
Taking the $\alpha \rightarrow \infty $ limit and using the finite $q$%
-binomial theorem \cite{Koe}: %\ref{eq:finiteqbinom}

\begin{equation}
(qx;q^{k})_{n}=\sum_{r=0}^{n}\frac{x^{r}}{[q]_{r}}b_{n,r}=%
\sum_{r=0}^{n}(-1)^{r}\QATOPD[ ] {r}{n}_{q^{k}}q^{{\frac{1}{2}}%
kr(r-1)+r}x^{r},
\end{equation}%
so that

\begin{equation}
{\frac{b_{n,r}}{[q]_{r}}}=(-1)^{r}\QATOPD[ ] {n}{r}_{q^{k}}q^{{\frac{1}{2}}%
kr(r-1)+r}.  \label{eq:bnr}
\end{equation}%
For later use note that one also have:

\begin{equation}
{\frac{b_{n,r}}{[q]_{r}}}=\frac{1}{r!}\left( \frac{d}{dx}\right)
^{(r)}(qx;q^{k})_{n}\rvert _{x=0}.
\end{equation}%
From this one gets:

\begin{equation}
\boxed{ Y_n(x,k|q) =
\frac{1}{[q]_n}\sum_{r=0}^{n}(-1)^r\qbin{n}{r}_{q^k}q^{{1\over2}r(r+1)+{1%
\over2}k r(r-1)}x^r.}  \label{eq:Y}
\end{equation}%
For $k=1$, both polynomials reduce to the Stieltjes-Wigert polynomials $%
\left( \ref{eq:SW}\right) $. Writing $Y_{n}(x,k|q)=y_{n,k}x^{n}+...$ and $%
Z_{n}(x,k|q)=z_{n,k}\;x^{nk}+...$ one finds:

\begin{equation}
z_{n,k}=\frac{(q^{-nk};q^{k})_{n}}{(q^{k};q^{k})_{n}^{2}}q^{\frac{1}{2}%
kn(kn-1)+kn(n+1)}=\frac{(-1)^{n}q^{\frac{1}{2}n^{2}k(k+1)}}{(q^{k};q^{k})_{n}%
},
\end{equation}
and 
\begin{equation}
y_{n,k}=\frac{1}{n![q]_{n}}q^{\frac{1}{2}n(n-1)}\left( \frac{d}{dx}\right)
^{(n)}(qx;q^{k})_{n}\rvert_{x=0}=\frac{(-1)^{n}q^{\frac{1}{2}(k+1)n(n-1)+n}}{%
[q]_{n}}.
\end{equation}
This leads to:

\begin{equation}
<Y_{n}(x,k|q),Z_{m}(x,k|q)>=h_{n}\delta _{n,m},
\end{equation}%
with respect to the measure $\frac{A\mbox{d}x}{[-x]_{\infty }[-q/x]_{\infty }%
},$ with $A$ such that $<1,1>=1$ and

\begin{equation}
h_{n}=\frac{q^{-nk}}{[q]_{n}}.
\end{equation}%
Using this, we can find for example:

\begin{equation}
Z^{P,Q}=N!(\frac{g_{s}}{2\pi })^{N/2}q^{-\frac{N}{2P^{2}}[-(1+\frac{1}{2}(1+%
\frac{P}{Q})(N-1))^{2}+1+\frac{4}{3}(N^{2}-1)]}\prod_{j=1}^{N-1}(1-q^{\frac{j%
}{PQ}})^{N-j},
\end{equation}%
which reduces to the known formula when $P=Q=1$ \cite{cs,Tierz}.

\section{Mathematical properties of the biorthogonal polynomials}

Since the biorthogonal Stieltjes-Wigert polynomials have not been addressed
in the literature, we derive here some of its fundamental properties.

\subsection{Behavior under dilatation}

First, we find some generating functions for $Z_{n}(x;k|q)$ and $%
Y_{n}(x;k|q) $ ($t\neq q^{-k}$)

\begin{equation}
\sum_{n\geq0}Z_{n}(x;k|q)t^{n}=\frac{f(tx^{k})}{(t;q^{k})_{\infty}},
\label{eq:P1Z}
\end{equation}
and

\begin{equation}
\sum_{n\geq0}\frac{[q]_{n}}{(q^{k};q^{k})_{n}}Y_{n}(x;k|q)t^{n}=\frac {g(tx)%
}{(t;q^{k})_{\infty}},  \label{eq:P1Y}
\end{equation}
with 
\begin{equation}
f(z)=\sum_{r\geq0}{\frac{q^{\frac{1}{2}j^{2}k(k+1)}}{(q^{k};q^{k})_{j}}}%
(-z)^{j}\text{ and }g(z)=\sum_{r\geq0}{\frac{q^{\frac{r(r-1)}{2}}}{r!}}%
z^{r}\left( \frac{d}{dx}\right) ^{(r)}(qx;q^{k})_{n}\rvert_{x=0}.
\end{equation}

We rely for this on formula $\left( 4.2\right) $ from \cite{AlSalam}. The
expression for $Z$ is essentially property $\left( 4.1\right) $ in \cite%
{AlSalam}. For $\left( \ref{eq:P1Y}\right) $ we use the explicit expression
obtained for $Y_{n}(x;k|q)$. Let us introduce the moment generating function:

\begin{equation}
G(t,x)\equiv\sum_{n\geq0}\frac{[q]_{n}}{(q^{k};q^{k})_{n}}Y_{n}(x;k|q)t^{n},
\end{equation}
that can be written as:

\begin{align}
G(t,x) & =\sum_{r\geq0}{\frac{q^{\frac{r(r-1)}{2}}}{r!}}x^{r}\left( \frac{d}{%
dx}\right) ^{(r)}\sum_{n\geq r}{\frac{(qx;q^{k})_{n}t^{n}}{(q^{k};q^{k})_{n}}%
}\rvert_{x=0} \\
& =\sum_{r\geq0}{\frac{q^{\frac{r(r-1)}{2}}}{r!}}x^{r}\left( \frac{d}{dx}%
\right) ^{(r)}\sum_{n\geq0}{\frac{(qx;q^{k})_{n}t^{n}}{(q^{k};q^{k})_{n}}}%
\rvert_{x=0}  \notag \\
& =\sum_{r\geq0}{\frac{q^{\frac{r(r-1)}{2}}}{r!}}x^{r}\left( \frac{d}{dx}%
\right) ^{(r)}{\frac{(qxt;q^{k})_{\infty}}{(t;q^{k})_{\infty}}}  \notag \\
& ={\frac{1}{(t;q^{k})_{\infty}}}\sum_{r\geq0}{\frac{q^{\frac{r(r-1)}{2}}}{r!%
}}(xt)^{r}\left( \frac{d}{dx}\right) ^{(r)}(qx;q^{k})_{n}\rvert
_{x=0}\quad;\quad t\neq q^{-k}  \notag
\end{align}

In the second line, the extra piece we add, being a degree $r-1$ polynomial
in $x$ does not contribute due to the derivative. In the third line we use
the $q$-binomial theorem, and in the fourth one we make the change of
variable $x\rightarrow xt$.

Now, by taking $\left( \ref{eq:P1Y}\right) $ with $x\rightarrow\lambda x,$
introducing in the r.h.s the factor ${\frac{(\lambda t;q^{k})_{\infty}}{%
(t;q^{k})_{\infty}}}\,$ and matching the coefficients of $t^{n}$ on both
sides one gets:

\begin{equation}
Y_{n}(\lambda x;k|q)=\sum_{j=0}^{n}\gamma_{nj}(\lambda)Y_{j}(x;k|q),
\end{equation}

\noindent with:

\begin{equation}
\boxed{ \gamma_{nj}(\lambda) =
{[q]_j\over[q]_n}{(q^k;q^k)_n\over(q^k;q^k)_j}{\lambda^j(\lambda;q^k)_{n-j}
\over(q^k;q^k)_{n-j}},}  \label{eq:dilatY}
\end{equation}

\noindent and similar steps involving Eq.~\ref{eq:P1Z} give:

\begin{equation}
Z_{n}(\lambda x;k|q)=\sum_{j=0}^{n}\zeta_{nj}(\lambda)Z_{j}(x;k|q),
\end{equation}

\noindent with

\begin{equation}
\boxed{
\zeta_{nj}(\lambda)={1\over(q^k;q^k)_{n-j}}\lambda^{kj}(%
\lambda^k;q^k)_{n-j}.}  \label{eq:dilatZ}
\end{equation}

One has $\zeta _{nj}=\gamma _{nj}$ for $k=1$ as it should. Moreover $\gamma
_{nn}=\zeta _{nn}=\lambda ^{n},$ by matching the dominant coefficients on
both sides. Note that equation (4.2) in \cite{AlSalam} contains a typo as it
does not fulfill this last condition (it would give $\zeta _{nn}=1$). And of
course one has $\zeta _{nj}(1)=\gamma _{nj}(1)=\delta _{nj}.$%.

Even though $\zeta_{nj}$ and $\gamma_{nj}$ are defined for $j\leq n$ we
extend for convenience their definition through

\begin{equation}
\zeta _{nj}=\gamma _{nj}=0\quad \text{if}\quad j>n.
\end{equation}

\subsection{Recurrence formulae}

The Stieltjes-Wigert polynomials $\left( \ref{eq:SW}\right)$ satisfy:

\begin{equation}
S_{n-1}(x|q)=(1-q^{n})S_{n}(x|q)+xq^{n}S_{n-1}(xq|q),  \label{eq:recSW}
\end{equation}%
an identity used by Chihara in \cite{Chi}, to prove that the zeros of the
polynomials satisfy:

\begin{equation}
x_{n,m}<x_{n-1,m}<qx_{n,m+1},
\end{equation}%
where $n$ denotes the order of the polynomial and $m$ indexes the zero. Note
that the zeros of the SW polynomials are an interesting quantity in the
context of topological strings \cite{Okuyama:2006eb}. In what follows, we
find the same identities for the biorthogonal polynomials.

\subsubsection{Fundamental recurrence relation}

Note that for the particular value $\lambda =q^{-1}$ $\left( \ref{eq:dilatZ}%
\right) $ gives: 
\begin{equation}
\zeta _{nj}(q^{-1})=0\quad \text{if}\quad j\leq n-2.
\end{equation}%
This implies the following simple recurrence relation for $Z_{n}(x,k|q)$:

\begin{equation}
\boxed{ Z_n(x,k|q)-Z_{n-1}(x,k|q)=q^{kn}Z_n(q^{-1}x,k|q).}  \label{314}
\end{equation}%
Certainly, if one writes the $Z_{n}(x,k|q)=\sum_{j=0}^{n}{\tau _{{n,j}}}%
x^{kj},$ it can be checked directly, from the explicit expression in $\left( %
\ref{eq:Z}\right) ,$ that one has: 
\begin{equation}
{\tau _{{n,j}}}-{\tau _{{n-1,j}}}=q^{k(n-j)}{\tau _{{n,j}},}
\end{equation}%
which implies the recurrence relation\footnote{%
Incidentally, this a check that Eq.~(\ref{eq:dilatZ}) is correct.}.

For $k=1,$ $\left( \ref{314}\right) $ reduces to the following relation for
the Stieltjes-Wigert polynomials

\begin{equation}
S_{n}(y)-S_{n-1}(y)=q^{n}S_{n}(q^{-1}y).  \label{eq:fundamS}
\end{equation}

\subsubsection{Moment generating recurrences}

From $\left( \ref{eq:bnr}\right) $ one has ($b_{n,0}=1$, $b_{n,-1}\equiv0$):

\begin{equation}
{\frac{b_{n+1,r}}{[q]_{r}}}={\frac{b_{n,r}}{[q]_{r}}}-q^{nk+1}{\frac{%
b_{n,r-1}}{[q]_{r-1}},}
\end{equation}%
which implies the following recurrence relation for the $Y_{n}(x,k|q):$

\begin{equation}
(1-q^{n+1})Y_{n+1}(x,k|q)=Y_{n}(x,k|q)-q^{nk+1}xY_{n}(qx,k|q),
\end{equation}%
or, equivalently:

\begin{equation}
\boxed{x
Y_n(x,k|q)=q^{-nk}\left(Y_n(q^{-1}x,k|q)-(1-q^{n+1})Y_{n+1}(q^{-1}x,k|q)%
\right).}  \label{eq:recY}
\end{equation}

We proceed in analogous way for $Z_{n}(x;k|q)$. For convenience we introduce
coefficients $c_{n,r},$ such that\footnote{%
One has $\tau_{{n,r}}={\frac{q^{{\frac{1}{2}}kr(kr-1)}}{(q^{k};q^{k})_{n}}}{%
\frac{c_{n,r}}{(q^{k};q^{k})_{r}}.}$}:

\begin{equation}
Z_{n}(x,k|q)\equiv\frac{1}{(q^{k};q^{k})_{n}}\sum_{r=0}^{n}\frac {x^{kr}q^{%
\frac{1}{2}kr(kr-1)}}{(q^{k};q^{k})_{r}}c_{n,r},
\end{equation}
that is:

\begin{equation}
{\frac{c_{n,r}}{(q^{k};q^{k})_{r}}}={\frac{(q^{-nk};q^{k})_{r}\ q^{kr(n+1)}}{%
(q^{k};q^{k})_{r}}.}
\end{equation}
Then, as in the previous case $c_{n,0}=1$, $c_{n,-1}\equiv0$, then:

\begin{equation}
{\frac{c_{n+1,r}}{(q^{k};q^{k})_{r}}}={\frac{c_{n,r}}{(q^{k};q^{k})_{r}}}%
-q^{nk+k}{\frac{c_{n,r-1}}{(q^{k};q^{k})_{r-1}},}
\end{equation}
and one gets the recurrence relation for $Z_{n}(x;k|q):$

\begin{equation}
(1-q^{k(n+1)})Z_{n+1}(x,k|q)=Z_{n}(x,k|q)-q^{nk+{\frac{k(k+1)}{2}}%
}x^{k}Z_{n}(q^{k}x,k|q),
\end{equation}
equivalently:

\begin{equation}
\boxed{x^k Z_n(x,k|q)=q^{-nk+{k(k-1)\over
2}}\left(Z_n(q^{-k}x,k|q)-(1-q^{k(n+1)})Z_{n+1}(q^{-k}x,k|q)\right).}
\label{eq:recZ}
\end{equation}%
One can easily check that these recurrence relations both reduce, taking $k=1
$, to $\left( \ref{eq:recSW}\right) .$

To conclude this Section, since we know the explicit behavior of the
polynomials under dilatation, we can employ $\left( \ref{eq:dilatY}\right) $
and $\left( \ref{eq:dilatZ}\right) $, and then, $\left( \ref{eq:recY}\right) 
$ and $\left( \ref{eq:recZ}\right) ,$ to obtain an explicit way to compute
the moments $<x^{l}Y_{n}(x,k|q)Z_{m}(x,k|q)>$. For instance, one has: 
\begin{align}
<xY_{n}(x,k|q)Z_{m}(x,k|q)>& =q^{-nk}k_{m}\left( \gamma
_{n,m}(q^{-1})-(1-q^{n+1})\gamma _{n+1,m}(q^{-1})\right) \\
& ={\frac{q^{-(n+m)k}}{[q]_{m}}}(\gamma _{n,m}(q^{-1})-(1-q^{n+1})\gamma
_{n+1,m}(q^{-1}).  \notag
\end{align}%
Note that this does not work so well when $k=1$ as $\gamma _{n,m}(q^{-1})$
is not well defined in this case, according to the discussion above.

\section{Stieltjes-Wigert, other results}

\subsection{Moment problem and $q$-2D Yang-Mills}

The Stieltjes-Wigert matrix model posses distinctive mathematical features,
in comparison with other, more usual models in the literature, such as
matrix models with Gaussian or polynomial potentials. The log-normal weight
function leads to an indeterminate moment problem \cite{Sti,moment} and
consequently, the Stieltjes-Wigert polynomials are not dense in $%
L^{2}(x,w\left( x\right) )$ (see \cite{Tierz,TdH1} for details). One of the
consequences, discussed in \cite{TdH1}, is the discretization of the
Chern-Simons matrix model. This opens the possibility for the discrete and
continuous versions of the same model to share the same orthogonal
polynomials. Note that in \cite{Arsiwalla:2005jb}, we find the suggestion of
studying the discrete matrix model (see below) with orthogonal polynomials,
as done in \cite{Gross:1994mr} for the 2D Yang-Mills theory case, that lead
to a discrete Gaussian matrix model. Recall that Gross and Matytsin \cite%
{Gross:1994mr} found a discrete version of the ordinary Gaussian matrix
model in their study of the $1/N$ expansion of the partition function of $2D$%
\ QCD on the sphere:

\begin{equation}
Z(A,N)\equiv \sum_{u_{1},\ldots ,u_{N}=-\infty }^{+\infty }\mathrm{e}^{-%
\frac{A}{2N}\sum_{i}u_{i}^{2}}\prod_{j<k}(u_{j}-u_{k})^{2}.
\end{equation}%
In contrast to the continuum case, that is solved with Hermite polynomials,
the discrete Gaussian weight does not have a closed system of orthogonal
polynomials associated (see \cite{Johansson}, for a recent discussion of
discrete matrix models). So, they could not rely on known orthogonal
polynomials, hence the difficulty of studying the discrete Gaussian model.
Actually, the large $N$ phase transition of the theory is related with the
discrepancy between the discrete and continuous orthogonal polynomials.

The orthogonal polynomials for the discrete matrix model in the Chern-Simons
case are the Stieltjes-Wigert polynomials, as we show below. Note that the
model in Chern-Simons/$q$-2D\ Yang-Mills differs from the Gaussian/2D
Yang-Mills model in the exponentiation of the eigenvalue repulsion at large
distances, $(u_{j}-u_{k})^{2}\rightarrow \sinh ^{2}\left( u_{i}-u_{j}\right)
.$ But this exponentiation is precisely the ultimate responsible of this
continuum/discrete equivalence (see \cite{TdH1}).

The following detailed computation highlights this special property of the
Chern-Simons matrix model. Let us proceed then to show the details of the
derivation from Eq. $\left( 25\right) $ to Eq. $\left( 26\right) $ in \cite%
{TdH1}

\begin{align}
Z_{d}& \equiv \sum_{u_{1},\ldots ,u_{N}=-\infty }^{+\infty }\mathrm{e}^{-{%
\frac{g_{s}}{2}}\sum_{i}u_{i}^{2}}\prod_{j<k}4\sinh ^{2}\left( {\frac{g_{s}}{%
2}}(u_{j}-u_{k})\right) \\
& =\sum_{u_{1},\ldots ,u_{N}=-\infty }^{+\infty }\mathrm{e}^{-{\frac{g_{s}}{2%
}}\sum_{i}u_{i}^{2}}\mathrm{e}^{(N-1)g_{s}\sum_{i}u_{i}}\prod_{j<k}\left(
e^{-g_{s}u_{j}}-e^{-g_{s}u_{k}}\right) ^{2}  \notag \\
& =\sum_{u_{1},\ldots ,u_{N}=-\infty }^{+\infty }q^{{\frac{1}{2}}%
\sum_{i}u_{i}^{2}}(cq)^{\sum_{i}u_{i}}\prod_{j<k}\left(
q^{u_{j}}-q^{u_{k}}\right) ^{2}  \notag \\
& =\sum_{u_{1},\ldots ,u_{N}=-\infty }^{+\infty
}\prod_{i=1}^{N}\,c^{u_{i}}\,q^{{\frac{1}{2}}u_{i}^{2}+u_{i}}\prod_{j<k}%
\left( q^{u_{j}}-q^{u_{k}}\right) ^{2}  \notag \\
& =c^{N(1-N)}\int \prod_{i=1}^{N}\mathrm{d}x_{i}\sum_{n=-\infty }^{+\infty
}c^{n}\,q^{{\frac{1}{2}}n^{2}+n}\delta (x-cq^{n})\prod_{j<k}\left(
x_{j}-x_{k}\right) ^{2}  \notag \\
& =c^{N(1-N)}q^{{\frac{N}{2}}}M(c)^{N}\int_{0}^{+\infty }\prod_{i=1}^{N}%
\mathrm{d}x_{i}w_{d}(x_{i})\prod_{j<k}\left( x_{j}-x_{k}\right) ^{2},  \notag
\end{align}

\noindent where we have introduced \footnote{%
Note that even though there are infinitely many discrete measures $w_{d}$
equivalent to $w$ (the parameter $c$ is a real number), in the Chern-Simons
case we are discussing, the constant $c$ is a function of $g_{s}$.} $c\equiv 
\mathrm{e}^{Ng_{s}}=q^{-N}$ and we recall that we have as usual $q\equiv 
\mathrm{e}^{-g_{s}}$. Moreover, one has:

\begin{equation}
w_{d}(x)\equiv {\frac{1}{\sqrt{q}\,M(c)}}\sum_{n=-\infty }^{+\infty
}\,c^{n}\,q^{{\frac{n^{2}}{2}}+n}\delta (x-cq^{n}),
\end{equation}

\noindent which is the discrete measure equivalent to the continuous
distribution $w(x)$ in $\left( \ref{SW}\right) $ as far as the integer
moments are concerned \cite{Chi,Chi2,Christ}. The normalization is given by:

\begin{equation}
M(c)\equiv(-cq^{3/2},-c^{-1}q^{-1/2},q;q)_{\infty}=[-cq^{3/2}]_{\infty
}[-c^{-1}q^{-1/2}]_{\infty}[q]_{\infty},
\end{equation}
therefore, we can use the equivalence between these two measures to write:

\begin{equation}
Z_{d}=c^{N(1-N)}q^{{\frac{N}{2}}}M(c)^{N}\int_{0}^{+\infty }\prod_{i=1}^{N}%
\mathrm{d}x_{i}\ w(x_{i})\ \prod_{j<k}\left( x_{j}-x_{k}\right) ^{2}.
\end{equation}%
However recall that, in the simplest case $P=Q=1$ in $\left( \ref{eq:ZPQ}%
\right) ,$ one has:

\begin{equation}
Z^{1,1}=q^{-{\frac{N^{3}}{2}}}\left( {\frac{g_{s}}{2\pi }}\right) ^{{\frac{N%
}{2}}}\int_{0}^{+\infty }\prod_{i=1}^{N}\mathrm{d}x_{i}\ w(x_{i})\
\prod_{j<k}\left( x_{j}-x_{k}\right) ^{2}.
\end{equation}%
Thus, one has a quite simple relation between the discrete and continuous
Stieltjes-Wigert ensembles:

\begin{equation}
\begin{split}
\left( \frac{g_{s}}{2\pi }\right) ^{-\frac{N}{2}}\int_{0}^{+\infty }&
\prod_{i}{\frac{\mbox{d}u_{i}}{2\pi }}e^{-\frac{u_{i}^{2}}{2g_{s}}%
}\prod_{i<j}\left( 2\sinh ({\frac{u_{i}-u_{j}}{2}})\right) ^{2}= \\
& \left( {\frac{q^{-{\frac{1}{2}}(1-2N+3N^{2})}}{[-q^{3/2-N}]_{\infty
}[-q^{N-1/2}]_{\infty }[q]_{\infty }}}\right) ^{N}\sum_{n_{1},\ldots
,n_{N}=-\infty }^{+\infty }e^{-{\frac{g_{s}}{2}}\sum_{i}n_{i}^{2}}%
\prod_{j<k}\left( 2\sinh \left( {\frac{g_{s}}{2}}(n_{j}-n_{k})\right)
\right) ^{2}
\end{split}%
\end{equation}%
Note the inversion of the coupling constant between the l.h.s. and r.h.s.

\subsection{From Stieltjes-Wigert to Rogers-Szeg\"{o}: unitary matrix model}

The Stieltjes-Wigert polynomials turn out to be intimately related to the
Rogers-Szeg\"{o} polynomials \cite{Carlitz,AN}, that are orthogonal on the
unit circle. This is useful to establish in detail the exact relationship
between the Stieltjes-Wigert matrix model and the unitary model considered
by Okuda \cite{Okuda:2004mb}. Following \cite{AN}, we recall the definition
and relations between the Rogers-Szeg\H{o} and Stieltjes-Wigert polynomial.
The Rogers-Szeg\H{o} polynomials are defined as:

\begin{equation}
H_{n}(z|q)\equiv \sum_{k=0}^{n}\QATOPD[ ] {n}{k}_{q}z^{k},
\end{equation}

\noindent and they satisfy an orthogonality relation on the complex unit
circle:

\begin{equation}
{\frac{1}{2i\pi}}\oint_{|w|=1}H_{m}(-q^{-1/2}\overline{w}\ |\
q)H_{n}(-q^{-1/2}w\ |\ q)\Theta_{3}\left( {\frac{\log w}{2i}}|\sqrt{q}%
\right) {\frac{\mathrm{d}w}{w}}={\frac{[q]_{m}}{q^{m}}}\delta_{mn},
\end{equation}

\noindent where $\Theta _{3}(z|q)$ is the third Jacobi theta function (see
definitions in the Appendix). Note that the orthogonality coefficients $%
h_{m}=\frac{[q]_{m}}{q^{m}}$ are identical to the ones (Stieltjes-Wigert)
that directly give the Chern-Simons partition function in the $S^{3}$ $U(N)$
case \cite{Tierz}. This is enough to write down an unitary matrix model for
the Chern-Simons partition function. However, let us show this point with
detail. The polynomials are also orthogonal with respect to a measure
defined on the full real line \cite{AN}:

\begin{equation}
{\frac{1} {\sqrt{\pi}}}\int_{-\infty}^{+\infty}H_{m}(-q^{-1/2}e^{-2i\mu x}\
|\ q)H_{n}(-q^{-1/2}e^{2i\mu x}\ |\ q)e^{-x^{2}}\mathrm{d}x={\frac{[q]_{m}}{%
q^{m}}}\delta_{mn},
\end{equation}

\noindent introducing $\mu$ through $q\equiv \mathrm{e}^{-2\mu^{2}}$. Now
consider the Stieltjes-Wigert polynomials $S_{n}\left( x\right) $ \cite%
{Szego}:

\begin{align}
S_{n}(x)& =\frac{(-1)^{n}q^{n/2+\frac{1}{4}}}{\sqrt{[q]_{n}}}\sum_{\nu
=0}^{n}\QATOPD[ ] {n}{\nu }_{q}q^{\nu ^{2}}(-\sqrt{q}x)^{\nu } \\
& =\frac{(-1)^{n}q^{n/2+\frac{1}{4}}}{\sqrt{[q]_{n}}}\hat{S}_{n}(-\sqrt{q}%
x|q),\text{ with }\hat{S}_{n}(z|q)\equiv \sum_{k=0}^{n}\QATOPD[ ] {n}{k}%
_{q}q^{k^{2}}z^{k}.  \notag
\end{align}%
These polynomials fulfill the following orthogonality relation on the real
line:

\begin{equation}
{\frac{1}{\sqrt{\pi }}}\int_{-\infty }^{+\infty }\hat{S}_{m}(-q^{-1/2}e^{-2%
\mu x}\ |\ q)\hat{S}_{n}(-q^{-1/2}e^{-2\mu x}\ |\ q)e^{-x^{2}}\mathrm{d}x={%
\frac{[q]_{m}}{q^{m}}}\delta _{mn}.
\end{equation}%
Using an elementary property of the $q$-binomial coefficients, the two
equivalent relationship follow:

\begin{equation}
H_{n}(x|q^{-1})=\hat{S}_{n}({q^{-n}x|q})\quad \text{and\quad }\hat{S}%
_{n}(x|q^{-1})=H_{n}({q^{-n}x|q}).
\end{equation}%
Then:

\begin{align}
<p_{n},p_{m}>_{w}& =\rho _{m,n}\frac{k}{\sqrt{\pi }}\int_{0}^{\infty }%
\mathrm{e}{^{-k^{2}\log ^{2}z}\hat{S}_{m}(-q^{1/2}z)\hat{S}_{n}(-q^{1/2}z)%
\mathrm{d}z} \\
& =\rho _{m,n}q^{-1/2}\frac{k}{\sqrt{\pi }}\int_{-\infty }^{\infty }\mathrm{e%
}{^{-k^{2}(x-{\frac{1}{2k^{2}}})^{2}}\hat{S}_{n}(-q^{1/2}\mathrm{e}^{x})\hat{%
S}_{n}(-q^{1/2}\mathrm{e}^{x})\mathrm{d}x}  \notag \\
& =\rho _{m,n}q^{-1/2}\frac{1}{\sqrt{\pi }}\int_{-\infty }^{+\infty }\mathrm{%
e}^{-x^{2}}\hat{S}_{n}(-q^{-1/2}\mathrm{e}^{\frac{-x}{k}})\hat{S}%
_{n}(-q^{-1/2}\mathrm{e}^{\frac{-x}{k}}){\mathrm{d}x}  \notag \\
& =\rho _{m,n}q^{-1/2}\frac{1}{\sqrt{\pi }}\int_{-\infty }^{+\infty }\mathrm{%
e}^{-y^{2}}H_{n}(-q^{-1/2}\mathrm{e}^{2i\mu x})H_{m}(-q^{-1/2}\mathrm{e}%
^{-2i\mu x})\mathrm{d}y  \notag \\
& =\rho _{m,n}q^{-1/2}{\frac{1}{2i\pi }}\oint_{|w|=1}H_{m}(-q^{-1/2}%
\overline{w}\ |\ q)H_{n}(-q^{-1/2}w\ |\ q)\Theta _{3}\left( {\frac{\log w}{2i%
}}\rvert \sqrt{q}\right) {\frac{\mathrm{d}w}{w}}  \notag \\
& =\rho _{m,n}q^{-1/2}\int_{0}^{2\pi }{\frac{\mathrm{d}\theta }{2\pi }}%
H_{m}(-q^{-1/2}\mathrm{e}^{-i\theta }\ |\ q)H_{n}(-q^{-1/2}\mathrm{e}%
^{i\theta }\ |\ q)\Theta _{3}\left( {\frac{\theta }{2}}\rvert \sqrt{q}%
\right) ,  \notag
\end{align}%
where we also have used, between lines $3$ and $4$, the fact that $\widehat{S%
}_{n}(a\mathrm{e}^{-2\kappa x}\ |\ q)$ and $H_{n}(a\mathrm{e}^{2i\kappa y}\
|\ q)$ are related by a Fourier transform \cite{AN}. We also introduced $%
2\mu =\frac{1}{\kappa}$ and: 
\begin{equation}
\rho_{m,n}=(-1)^{m+n}\frac{q^{\frac{m+n+1}{2}}}{ \sqrt{[q]_{m}[q]_{n}}}.
\end{equation}

%Between line 3 and 4 we make use of Eq.~%
%\ref{eq:wvfction} and the Parseval identity and also use the fact that $q\in
%\mathbb{R}$.
The next line is given by the results of the previous section. Now consider
Eq.~(3.22) in \cite{Okuda:2004mb}, it reads:

\begin{align}
\tilde{Z}_{CS}& ={\frac{1}{|W|}}\int \left( \prod_{i=1}^{N}{\frac{\mathrm{d}%
\theta _{i}}{2\pi }}\Theta _{00}(e^{i\theta _{i}}|q)\right)
\prod_{i<j}\left( \sin ({\frac{\theta _{i}-\theta _{j}}{2}})\right) ^{2} \\
& ={\frac{(-1)^{\frac{N(N-1)}{2}}}{|W|}}\int \left( \prod_{i=1}^{N}{\frac{%
\mathrm{d}\theta _{i}}{2\pi }}\Theta _{00}(e^{i\theta _{i}}|q)\right)
\prod_{i<j}(e^{i\theta _{i}}-e^{i\theta _{j}})\prod_{i<j}(e^{-i\theta
_{i}}-e^{-i\theta _{j}})  \notag \\
& ={\frac{(-1)^{\frac{N(N-1)}{2}}}{|W|}}\int \left( \prod_{i=1}^{N}{\frac{%
\mathrm{d}\theta _{i}}{2\pi }}\Theta _{00}(e^{i\theta _{i}}|q)\right)
\det_{1\leq i,j\leq N}(H_{j-1}(e^{i\theta _{i}}))\det_{1\leq i,j\leq
N}(H_{j-1}(e^{-i\theta _{i}})),  \notag
\end{align}%
where 
\begin{equation}
\Theta _{00}(e^{i\theta }|q)=\sum_{j\in \mathbb{Z}}q^{{\frac{j^{2}}{2}}%
}e^{ij\theta }.
\end{equation}%
Then, considering that \cite{Okuda:2004mb} and \cite{AN} have different
conventions for the third Jacobi function one sees that: 
\begin{equation}
\Theta _{00}^{(O)}\left( e^{i\theta }|q\right) =\Theta _{3}^{(A)}\left( {%
\frac{\theta }{2}}|\sqrt{q}\right) .
\end{equation}%
Therefore, one can continue the computation and write:

\begin{equation}
\tilde{Z}_{CS}={\frac{(-1)^{\frac{N(N-1)}{2}}N!}{|W|}}<\left( \det_{1\leq
i,j\leq N}((-1)^{j-1}q^{-{\frac{j-1}{2}}}\sqrt{[q]_{j-1}}p_{j-1}(z_{i}))%
\right) ^{2}>_{w},
\end{equation}%
which then connects with the usual expression of the partition function in
terms of the orthogonal polynomials for the measure on the real line.

%and $\tilde{H}_{i}$ are the corresponding orthogonal and monic polynomials.
%Note that this doesn't seem to be quite the same as $H_{i}$ as there's some
%annoying $2\pi $ factor in the theta-function measure. Maybe I've done some
%mistake in some conventions...

Incidentally, both Stieltjes-Wigert and Rogers-Szeg\"{o} can be interpreted
as the ground-state wavefunction of a $q$-deformed harmonic oscillator \cite%
{AN}. This is an appealing property as it has been recently shown that the
Stieltjes-Wigert polynomial describes $B$-brane amplitudes on the conifold 
\cite{Okuyama:2006eb}.

\subsection{Quantum dimensions as averages of Schur polynomials in the
Stieltjes-Wigert ensemble}

\bigskip

In this section we prove a formula for the averages of Schur polynomials
that appears in \cite{Marino:2004eq,hyunyi}, without relying on the
equivalence with Chern-Simons theory.

Recall that Schur polynomials $\mathfrak{s}_{\lambda }$ \cite{macdonald}
constitute a basis of symmetric functions in a given set of variables $%
x=(x_{i})$ and are indexed by Young diagrams $\lambda $. If the variables $x$
are seen as eigenvalues of some matrix $M\in sl_{n}$ then $\mathfrak{s}%
_{\lambda }(M)\equiv \mathrm{Tr}_{\lambda }(M)$ is the trace of $M$ in the
representation associated to $\lambda $. The Schur polynomials may also be
more directly defined in terms of the skew-symmetric polynomials $\mathfrak{a%
}_{\mu }=\det (x_{i}^{\mu _{j}+n-j})$ as: 
\begin{equation}
\mathfrak{s}_{\lambda }(x)\equiv \frac{\mathfrak{a}_{\lambda +\delta }(x)}{%
\mathfrak{a}_{\delta }(x)},
\end{equation}%
where $\mathfrak{a}_{\delta }(x)$ is the Vandermonde in the variables $x$.
The result we want to show is the following:

\begin{equation}
<\mathfrak{s}_{\lambda }(M)>_{w}=q^{-n|\lambda |-\frac{1}{2}C_{\lambda
}^{U(n)}}\mathcal{D}_{\lambda }.  \label{eq:schuraverage}
\end{equation}%
with 
\begin{equation}
C_{\lambda }^{U(n)}=(n+1)|\lambda |+\sum_{i}(\lambda _{i}^{2}-2i\lambda
_{i}),  \label{eq:casimir}
\end{equation}%
the Casimir of the representation labeled by the Young diagram $\lambda $
and $|\lambda |$ its total number of boxes. Background material for this
Section is presented in Appendix \ref{s:determinant}. Note that this
quantity can be rewritten using Eq.$\left( \ref{eq:nnumber}\right) $ as: 
\begin{equation}
C_{\lambda }^{U(n)}=n|\lambda |+2(n(\lambda ^{\prime })-n(\lambda )),
\end{equation}%
where $\lambda ^{\prime }$ denotes the conjugate partition. The quantum
dimension is defined by the $q$-hook formula\footnote{%
See for instance \S 4.4 in \cite{fuchs} for a clear discussion of the
definition and properties of quantum dimensions.} 
\begin{equation}
\mathcal{D}_{\lambda }\equiv \prod_{x\in \lambda }\frac{\lfloor
n+c(x)\rfloor }{\lfloor h(x)\rfloor }.  \label{eq:qdimension}
\end{equation}%
where for each box $x=(i,j)$ of the diagram $h(x)\equiv \lambda _{i}+\lambda
_{j}^{\prime }-i-j+1$ is the hook-length and $c(x)\equiv j-i$ the content of 
$x$.

\subsubsection{Case of 1-column diagrams}

%%%%%%%%%%%%%%%%%%%%%%%%%%%%%%%%%%%%%%%%%%%%%%%%%%
The quantum dimension of the j-th fundamental representation of $sl_{n}$,
which is associated\footnote{%
We will freely switch notations between Young diagrams and partitions in the
following.} to the partition $(1^j)$, or a one-column Young tableau of
length $j$, is

\begin{equation}
\mathcal{D}_{(1^{j})}=\dim _{q}\Lambda _{(j)}=\QATOPD\lfloor \rfloor
{n}{j}_{q}.
\end{equation}%
Moreover, the monic Stieltjes-Wigert polynomials can be written\footnote{%
The measure being here normalized such that $<1>_{w}=1$.}:

\begin{equation}
\pi _{n}(x)=\sum_{j=0}^{n}(-1)^{n-j}q^{(j-n)(j+n+\frac{1}{2})}\QATOPD[ ] {n}{%
j}_{q}x^{j}=<\det (x-M)>_{w}.
\end{equation}%
Besides, the following formula holds for the characteristic polynomial:

\begin{equation}
\det (x-M)=\sum_{j=0}^{n}(-1)^{n-j}\mathfrak{s}_{(1^{n-j})}(M)x^{j},
\label{eq:characteristicpol}
\end{equation}%
with $\mathfrak{s}_{\lambda }(M)$ the Schur polynomial associated to the
partition $\lambda $. Therefore: 
\begin{equation}
\sum_{j=0}^{n}(-1)^{n-j}<\mathfrak{s}_{(1^{n-j})}(M)>_{w}x^{j}=%
\sum_{j=0}^{n}(-1)^{n-j}q^{(j-n)(j+n+\frac{1}{2})}\QATOPD[ ] {n}{j}_{q}x^{j},
\end{equation}%
from which we extract:

\begin{equation}
\boxed{ <\mathfrak{s}_{(1^j)}(M)>_w = q^{-j(2n-j+\frac{1}{2})}\qbin{n}{j}_q
= q^{-\frac{j}{2}(3n-j+1)}\qfloor{n}{j}_q .}  \label{eq:schurfundam}
\end{equation}%
%
%
%For later convenience let's also write
%$<\mathfrak{s}_{(1^{n-j})}(M)>=q^{(j-n)(n+{j+1\over2})}\mathcal{D}_{(1^{n-j})}$
Using $\left( \ref{eq:casimir}\right) $, one sees that Eq.~$\left( \ref%
{eq:schurfundam}\right) $ is indeed consistent with Eq.~$\left( \ref%
{eq:schuraverage}\right) $.

%%%%%%%%%%%%%%%%%%%%%%%%%%%%%%%%%%%%%%%%%%%%%%%%%%%%%%%%%%%%%%%%

\subsubsection{General case}

%%%%%%%%%%%%%%%%%%%%%%%%%%%%%%%%%%%%%%%%%%%%%%%%%%%%%%%%%%%%%%%%
\label{s:general case}

To study the case of a general Young diagram we note that as a
generalization of Eq.~\ref{eq:characteristicpol}, higher powers of the
characteristic polynomial are generating functions for diagrams with a
higher number of columns. Relying on a formula first computed in \cite{BH}
we then relate the average of Schur polynomials to some determinant of
Stieltjes-Wigert polynomials. From \cite{macdonald} (I.4 Example 5 p.67) we
see (taking a slightly more convenient notation)

\begin{equation}
\prod_{i=1}^{k}\prod_{j=1}^{n}(x_{i}+y_{j})=\sum_{\lambda \,;\,\lambda
_{1}\leq k}\mathfrak{s}_{\lambda }(y)\mathfrak{s}_{\tilde{\lambda}^{\prime
}}(x),
\end{equation}%
where $\lambda =(\lambda _{1},\ldots ,\lambda _{n})$ is a Young diagram with
at most $k$ columns as imposed by the condition $\lambda _{1}\leq k$. The
associate diagram $\tilde{\lambda}^{\prime }$ is defined as $(n-\lambda
_{n},\ldots ,n-\lambda _{1})$.

Therefore, if one considers $-y_{j}$ to be the eigenvalues of $M,$ one
immediately gets:

\begin{equation}
\boxed{ \prod_{i=1}^k\det(x_i-M)=\sum_{\lambda\,;\,\lambda_1\leq
k}(-1)^{|\lambda|}\mathfrak{s}_{\lambda}(M)\mathfrak{s}_{\tilde{%
\lambda}'}(x) .}  \label{eq:lhs}
\end{equation}

By a standard result on characteristic polynomials \cite{BH}, we have: 
\begin{equation}
<\prod_{i=1}^{k}\det (x_{i}-M)>_{w}=\frac{1}{\mathfrak{a}_{\delta }(x)}%
\left\vert 
\begin{matrix}
\pi _{n}(x_{1}) & \ldots & \pi _{n+k-1}(x_{1}) \\ 
\vdots & \vdots & \vdots \\ 
\pi _{n}(x_{k}) & \ldots & \pi _{n+k-1}(x_{k})%
\end{matrix}%
\right\vert ,
\end{equation}%
with $a_{\delta }(x)$ the Vandermonde determinant of the $x$ variables.

We now turn to the r.h.s., that we call $\Delta $ for convenience. From the
explicit expression for the Stieltjes-Wigert polynomials one obtains:

\begin{eqnarray*}
\mathfrak{a}_{\delta }(x)\Delta &=&\sum_{i_{1},\ldots ,i_{k}}\sum_{\sigma
\in \mathfrak{S}_{k}}\epsilon (\sigma )(-1)^{i_{1}+\ldots
+i_{k}}\prod_{j=1}^{k}q^{-i_{j}(2n+2\sigma (j)-2-i_{j}+\frac{1}{2}%
)}\prod_{j=1}^{k}\QATOPD[ ] {n+\sigma (j)-1}{i_{j}}x_{j}^{n+\sigma
(j)-1-i_{j}} \\
&=&\sum_{i_{1},\ldots ,i_{k}}(-1)^{i_{1}+\ldots +i_{k}+\frac{k(k-1)}{2}%
}\prod_{j=1}^{k}q^{-i_{j}(2n-i_{j}+\frac{1}{2})-(j-1)(2n+j-\frac{1}{2})} \\
&&\times \left( \sum_{\sigma \in \mathfrak{S}_{k}}\epsilon (\sigma
)\prod_{j=1}^{k}\QATOPD[ ] {n+\sigma (j)-1}{i_{j}+\sigma (j)-1}\right)
\prod_{j=1}^{k}x_{j}^{n-i_{j}} \\
&=&\sum_{i_{1},\ldots ,i_{k}}(-1)^{i_{1}+\ldots +i_{k}+\frac{k(k-1)}{2}%
}\prod_{j=1}^{k}q^{-i_{j}(2n-i_{j}+\frac{1}{2})-(j-1)(2n+j-\frac{1}{2})} \\
&&\times \det_{1\leq a,b\leq k}\left( \QATOPD[ ] {n+b-1}{i_{a}+b-1}\right)
\prod_{j=1}^{k}x_{j}^{n-i_{j}},
\end{eqnarray*}%
in the second line we have relabeled $i_{j}\rightarrow i_{j}+\sigma (j)-1$.
Now we study the determinant in the previous expression and show that if $%
i_{1}>\ldots >i_{n}:$

%\begin{equation}
%\det_{1\leq a,b\leq k}\left(\qbin{n+b-1}{i_a+b-1}\right)
%\end{equation}

\begin{equation}
\boxed{ \det_{1\leq a,b\leq
k}\left(\qbin{n+b-1}{i_a+b-1}\right)=A_q(\lambda)\qbin{n}{\lambda},}
\label{eq:determinant}
\end{equation}%
with the constant $A_{q}(\lambda )=q^{n(\lambda ^{\prime })},$ as we show in
appendix \ref{s:determinant}. 
% and $\o\in\mathfrak{S}_k$ such that $\o(i_1)>\ldots>\o(i_k)$.
Here $\lambda ^{\prime }$ is the partition conjugate to $\lambda $ and equal
to $(i_{1},i_{2}+1,\ldots ,i_{k}+k-1)$. $\QATOPD[ ] {n}{\lambda }$ is a
notation generalizing the $q$-binomial coefficients $\QATOPD[ ] {n}{j}$ and
which is defined by the $q$-hook formula\footnote{%
We warn the reader that for convenience we adopt a slightly different
notation for $\QATOPD[ ] {n}{\lambda }$ compared to \cite{macdonald} in the
sense that its value for the partition $(1^{r})$ is the usual Gaussian
polynomial $\QATOPD[ ] {n}{r}$ whereas in \cite{macdonald}, $\QATOPD[ ] {n}{r%
}=\QATOPD[ ] {n}{(r)}$.}

\begin{equation}
\QATOPD[ ] {n}{\lambda }\equiv \prod_{x\in \lambda }\frac{1-q^{n+c(x)}}{%
1-q^{h(x)}},
\end{equation}%
This is nothing but the analog of quantum dimension Eq.~$\left( \ref%
{eq:qdimension}\right) $ where instead of using the $\lfloor .\rfloor $
version of the $q$-integers one rather uses\footnote{%
Recall that $[n]=q^{\frac{n-1}{2}}\lfloor n\rfloor $} $[.]$. Identifying
with the l.h.s. of Eq.~$\left( \ref{eq:lhs}\right) $ we obtain: 
\begin{equation}
<\mathfrak{s}_{\lambda }(M)>_{w}=q^{\sum_{j}-i_{j}(2n-i_{j}+\frac{1}{2}%
)-(j-1)(2n+j-\frac{1}{2})}q^{n(\lambda ^{\prime })}\QATOPD[ ] {n}{\lambda }.
\end{equation}%
The last step we need to perform now is to convert $\QATOPD[ ] {n}{\lambda }$
in terms of $\mathcal{D}_{\lambda }$ and check that the prefactor is given
by Eq.~$\left( \ref{eq:casimir}\right) $. To this end we first note that due
to Eq.~$\left( \ref{eq:hooklength}\right) $ and Eq.~$\left( \ref{eq:content}%
\right) $ we have:

\begin{equation}
\QATOPD[ ] {n}{\lambda }=q^{\frac{1}{2}(n-1)|\lambda |-n(\lambda )}\mathcal{D%
}_{\lambda },
\end{equation}%
and 
\begin{equation}
\sum_{j}-i_{j}(2n-i_{j}+\frac{1}{2})-(j-1)(2n+j-\frac{1}{2})=-(2n-\frac{3}{2}%
)|\lambda |+\sum_{j}\lambda _{j}^{^{\prime }2}-2j\lambda _{j}^{\prime }.
\end{equation}%
To rewrite things in terms of the partition itself, rather than its
transposed, we use the relationship: 
\begin{equation}
\sum_{i}\lambda _{i}^{2}-2i\lambda _{i}=2(n(\lambda ^{\prime })-n(\lambda
))-|\lambda |.
\end{equation}%
Collecting all the prefactors we can eventually write our final result: 
\begin{equation}
\boxed{
<\mathfrak{s}_\lambda>=q^{-\frac{1}{2}\left((3n+1)|\lambda|+\sum_i%
\lambda_i^2-2i\lambda_i\right)}\mathcal{D}_\lambda ,}
\end{equation}%
which coincides with Eq.~$\left( \ref{eq:schuraverage}\right) $.

\section{Conclusions and Outlook}

We have constructed the biorthogonal Stieltjes-Wigert polynomials, necessary
for computing expressions such as $\left( \ref{lens}\right) $, that appear
in Chern-Simons theory. The polynomials are not discussed in the mathematics
literature, so a great deal of the effort has been devoted to the explicit
description of their fundamental properties.

The construction of the biorthogonal Stieltjes-Wigert polynomials may very
well be a necessary technical step for the computation of knot invariants in
exact fashion employing orthogonal polynomials. Note that, so far, the
topological invariants computed with orthogonal polynomials are only
Chern-Simons partition functions (more precisely, only the case of $S^{3}$
with gauge group $U(N)$ \cite{Tierz}). Indeed, according to Mari\~{n}o \cite%
{comm}, the results in \cite{Marino:2002fk} can be extended in order to
obtain random matrix descriptions of other Chern-Simons observables. The
case of torus knots for example, amounts to:

\begin{eqnarray}
W_{R}^{(P,Q)} &=&{\frac{\mathrm{e}^{-{\frac{g_{s}}{2}}\bigl(PQ(\Lambda
^{2}-\rho ^{2})+({\frac{P}{Q}}+{\frac{Q}{P}})\rho ^{2}\bigr)}}{|PQ|^{\frac{N%
}{2}}N!}}  \notag \\
&&\times \int \prod_{i=1}^{N}{\frac{du_{i}}{2\pi }}\mathrm{e}%
^{-\sum_{i}u_{i}^{2}/2g_{s}}\prod_{i<j}\left( 2\sinh {\frac{u_{i}-u_{j}}{2P}}%
\right) \left( 2\sinh {\frac{u_{i}-u_{j}}{2Q}}\right) S_{\lambda }(\mathrm{e}%
^{u_{i}}),  \label{gen}
\end{eqnarray}%
where $S_{\lambda }(x_{i})$ are Schur polynomials associated to the
partition $\lambda $ (representations of $U(N)$ are labelled by partitions $%
\lambda )$.

That is to say, the $\left( \ref{lens}\right) $ solved here, but with an
insertion of a Schur polynomial. However, an obstacle can be the lack of a
computational device for random matrix-like quantities with such a term.
Note that the case of an ordinary Gaussian Hermitian matrix model with a
Schur polynomial insertion, was solved in \cite{DiFrancesco:1992cn} by
purely combinatorial methods, with no use of Hermite polynomials at all.
Nevertheless, as we have seen in the last Section, a computation of the
Stieltjes-Wigert ensemble with the Schur polynomial can be carried out with
a mixture of combinatorial and orthogonal polynomials techniques. Therefore,
it turns out that we have studied in detail the two cases comprised in $%
\left( \ref{gen}\right) .$ The biorthogonal case without the Schur insertion
and the average of the Schur polynomial in the orthogonal ($P=Q=1$) ensemble.

We have also studied other aspects of the (ordinary) Stieltjes-Wigert
polynomials that are of direct relevance for the corresponding matrix model.
In particular, we have discussed in detail the equivalence of the
Chern-Simons matrix model with its discrete counterpart, of very much
interest in $q$-$2D$ Yang-Mills theory, and also discussed the close ties
with Rogers-Szeg\"{o} polynomials, which are defined on the unit circle
(both sets of polynomials being an equivalent solution of a $q$-deformed
harmonic oscillator problem). This relationship allows to clearly establish
the relationship with the unitary matrix model discussed in \cite%
{Okuda:2004mb}. Fundamental properties of the Stieltjes-Wigert polynomials
like their asymptotic behavior and the above mentioned $q$-deformed harmonic
oscillator property, may be of interest in connection with the recently
established role of the polynomial in the study of topological strings \cite%
{Okuyama:2006eb}. We hope to address some of these issues in future work.

\bigskip

\textbf{Acknowledgments}

\bigskip

We are grateful to Edouard Br\'{e}zin, Bernard Julia, Vladimir Kazakov and
Marcos Mari\~{n}o for valuable comments and correspondence.

\newpage

\appendix

\section{Normalizations}

To follow standard conventions it is convenient to have the orthogonal
polynomials either monic or normalized to unity. Hence, we rewrite the
previous ones a little bit (it will also make the link with the usual
Stieltjes-Wigert for $k=1$ more transparent).

\subsection{Notations, $k=1$}

From Szeg\"o \cite{Szego} we have for the Stieltjes-Wigert polynomials

\begin{equation}
p_{n}(x)=\frac{(-1)^{n}q^{n/2+\frac{1}{4}}}{\sqrt{[q]_{n}}}\sum_{\nu =0}^{n}%
\QATOPD[ ] {n}{\nu }_{q}q^{\nu ^{2}}(-\sqrt{q}x)^{\nu },
\end{equation}%
with $\QATOPD[ ] {n}{\nu }_{q}$ the $q$-binomial coefficient:

\begin{equation*}
\QATOPD[ ] {n}{\nu }_{q}\equiv \frac{\lbrack q]_{n}}{[q]_{\nu }[q]_{n-\nu }}.
\end{equation*}%
These polynomials are orthonormal for the scalar product $<,>_{w}$ induced
by: 
\begin{equation}
w(x)=\frac{\kappa }{\sqrt{\pi }}\ e^{-\kappa ^{2}\log ^{2}x},
\label{eq:continuous}
\end{equation}%
with $q=e^{-1/2\kappa ^{2}}$ as usual. Note that one has:

\begin{equation}
<1,1>_w=1/\sqrt{q}.
\end{equation}
The $S_n$ polynomials in Eq.~(\ref{eq:SW}) to which Askey refers \cite{Askey}
as the SW polynomials are written in a slightly different form. They satisfy

\begin{equation}
<S_n,S_m>=\frac{q^{-n}}{[q]_n}\delta_{n,m},
\end{equation}
with $<,>$ the scalar product associated to the measure $\frac{A\mbox{d} x}{%
[-x]_\infty[-q/x]_\infty}$ with $A$ a normalization constant such that $%
<1,1>=1$.

Then, the polynomials defined by: 
\begin{equation}
\tilde{S}_{n}(x)\equiv (-1)^{n}\sqrt{[q]_{n}}q^{n/2}S_{n}(x),
\end{equation}%
are orthonormal for $<,>$. One then sees that:

\begin{equation}
p_{n}(x)=q^{1/4}\tilde{S}_{n}(\sqrt{q}x).
\end{equation}%
Since Al-Salam and Verma \cite{AlSalam} have the same notations as Askey for
the $q$-Laguerre polynomials, we will make the same rewriting when using the
biorthogonal Stieltjes-Wigert polynomials in the context of Chern-Simons
theory computations.

\subsection{$k$ arbitrary}

Normalizing and changing variables as in the previous section we define new
polynomials

\begin{equation}
\boxed{R_n(x,k|q)\equiv\frac{(-1)^nq^{1/4}}{\sqrt{k_n}}Y_n(%
\sqrt{q}x,k|q)=r_{n,k}x^n+...}
\end{equation}
and

\begin{equation}
\boxed{
T_n(x,k|q)\equiv\frac{(-1)^nq^{1/4}}{\sqrt{k_n}}Z_n(%
\sqrt{q}x,k|q)=t_{n,k}x^{nk}+...}
\end{equation}
Then one has

\begin{equation}
r_{n,k}=\frac{q^{(n+\frac{1}{2})^2}}{\sqrt{[q]_n}}q^{\frac{(k-1)n^2}{2}},
\end{equation}
and

\begin{equation}
t_{n,k}=\frac{\sqrt{[q]_{n}}}{(q^{k};q^{k})_{n}}q^{(nk+\frac{1}{2})^{2}-%
\frac{1}{2}n^{2}k(k-1)},
\end{equation}%
which reduce to $\frac{q^{(n+1/2)^{2}}}{\sqrt{[q]_{n}}}$ when $k=1$ as
expected. Then one has

\begin{equation}
<R_n(x,k|q) T_n(x,k|q)>_w=\delta_{m,n}.
\end{equation}
Following Eq.~(\ref{eq:recY}) and (\ref{eq:recZ}) the recurrence relations
for these orthonormal polynomials read

\begin{equation}
\boxed{x
R_n(x,k|q)=q^{-nk-1/2}\left(R_n(q^{-1}x,k|q)+q^{-k/2}%
\sqrt{1-q^{n+1}}R_{n+1}(q^{-1}x,k|q)\right),}  \label{eq:recR}
\end{equation}

\begin{equation}
\boxed{x^k T_n(x,k|q)=q^{-nk+{k(k-2)\over
2}}\left(T_n(q^{-k}x,k|q)+q^{-k/2}(1-q^{k(n+1)})\sqrt{k_{n+1}\over
k_n}T_{n+1}(q^{-k}x,k|q)\right).}  \label{eq:recT}
\end{equation}

\section{Proof of Eq.~$\left( \protect\ref{eq:determinant}\right) $}

\label{s:determinant}

Eq.~$\left( \ref{eq:determinant}\right) $ is not obvious at first sight
because when $q=1$, Giambelli's formula\footnote{%
See for instance \cite{difrancescoQFT} Eq.~(16.114) for a nice presentation.}
would lead us to write:

\begin{equation}
\det_{1\leq a,b\leq k}\left( \binom{n}{i_{a}+b-1}\right) =\dim \lambda .
\end{equation}%
However, classically one also has:

\begin{equation}
\det_{1\leq a,b\leq k}\left( \binom{n}{i_{a}+b-1}\right) =\det_{1\leq
a,b\leq k}\left( \binom{n+b-1}{i_{a}+b-1}\right) ,
\end{equation}%
which can be seen to hold by using Pascal's identity. Nevertheless, Pascal's
identity in the quantum case, is slightly more complicated and reads:

\begin{equation}
\QATOPD[ ] {n+1}{j+1}=\QATOPD[ ] {n}{j+1}+q^{n-j}\QATOPD[ ] {n}{j},
\label{eq:qpascal}
\end{equation}%
and we thus see that in the quantum case the same kind of simplification can
not be shown to hold as simply as in the classical setting.

To prove Eq.~$\left( \ref{eq:determinant}\right) $ nevertheless, first
recall that in the space of symmetric polynomials the change of basis
between the elementary symmetric polynomials\footnote{%
Elementary symmetric polynomials are special cases of Schur polynomials of
1-column diagrams, or $e_{r}=\mathfrak{s}_{\Lambda _{r}}$ in our notations.} 
$e_{r}$ and the Schur polynomials $\mathfrak{s}_{\lambda }$ is given by (%
\cite{macdonald} Eq.~(3.5)):

\begin{equation}
\mathfrak{s}_{\lambda }=\det (e_{\lambda _{i}^{\prime }-i+j}),
\label{eq:basischange}
\end{equation}%
where $\lambda =(\lambda _{1}\geq \ldots \geq l_{n})$ is a partition and $%
\lambda ^{\prime }$ its conjugate partition. This is actually nothing else
than Giambelli's identity, written for the symmetric polynomials and not
just for the dimensions of the associated representations. To compute from
this, note that if one considers $x=(1,q,\ldots ,q^{n-1}),$ then one has (%
\cite{macdonald} \S I.3 Ex.~1. p.44): 
\begin{equation}
\mathfrak{s}_{\lambda }(x)=q^{n(\lambda )}\QATOPD[ ] {n}{\lambda },
\end{equation}%
where $n(\lambda )$ is defined as: 
\begin{equation}
n(\lambda )\equiv \sum_{i\geq 1}(i-1)\lambda _{i}=\sum_{j\geq 1}\binom{%
\lambda ^{\prime }}{2},  \label{eq:nnumber}
\end{equation}%
and satisfies the following useful formulae:

\begin{equation}
\sum_{x\in \lambda }c(x)=n(\lambda ^{\prime })-n(\lambda )
\label{eq:content}
\end{equation}%
for the content (\cite{macdonald}, \S 1 Ex.3, p11 ), and another one for the
hook-lengths (\cite{macdonald}, \S 1 Ex.2, p.11 ): 
\begin{equation}
\sum_{x\in \lambda }h(x)=n(\lambda )+n(\lambda ^{\prime })+|\lambda |.
\label{eq:hooklength}
\end{equation}%
. Let us come back to our computation and particularize Eq.~$\left( \ref%
{eq:basischange}\right) $ to $x=(1,q,\ldots ,q^{n-1}),$ that gives:

\begin{equation}
q^{n(\lambda )}\QATOPD[ ] {n}{\lambda }=\det \left( q^{\frac{(\lambda
_{i}^{\prime }-i+j)(\lambda _{i}^{\prime }-i+j-1)}{2}}\QATOPD[ ] {n}{\lambda
_{i}^{\prime }-i+j}\right) ,
\end{equation}%
or, introducing $i_{a}+a-1=\lambda _{a}^{\prime }$,

\begin{equation}
q^{n(\lambda )}\QATOPD[ ] {n}{\lambda }=\det \left( q^{\frac{%
(i_{a}+b-1)(i_{a}+b-2)}{2}}\QATOPD[ ] {n}{i_{a}+b-1}\right) .
\end{equation}%
This is still not quite what we want. To proceed further note that according
to Eq.~$\left( \ref{eq:qpascal}\right) $ one has:

\begin{equation}
q^{\frac{j(j+1)}{2}}\QATOPD[ ] {n+1}{j+1}=q^{\frac{j(j+1)}{2}}\QATOPD[ ] {n}{%
j+1}+q^{n}\left( q^{\frac{j(j-1)}{2}}\QATOPD[ ] {n}{j}\right) .
\end{equation}%
Therefore, by multiple linear combinations of columns one can write:

\begin{equation}
\det \left( q^{\frac{(i_{a}+b-1)(i_{a}+b-2)}{2}}\QATOPD[ ] {n}{i_{a}+b-1}%
\right) =\det \left( q^{\frac{(i_{a}+b-1)(i_{a}+b-2)}{2}}\QATOPD[ ] {n+b-1}{%
i_{a}+b-1}\right) .
\end{equation}%
Now, for convenience extract a factor $q^{\frac{i_{a}(i_{a}-1)}{2}}$ in each
line to get:

\begin{equation}
\det \left( q^{\frac{(i_{a}+b-1)(i_{a}+b-2)}{2}}\QATOPD[ ] {n+b-1}{i_{a}+b-1}%
\right) =q^{\sum_{a}{\frac{i_{a}(i_{a}-1)}{2}}}\det \left( q^{\frac{%
(b-1)(2i_{a}+b-2)}{2}}\QATOPD[ ] {n+b-1}{i_{a}+b-1}\right) .
\end{equation}%
To proceed further note the following property of the $q$-binomial
coefficients: 
\begin{equation}
q^{i_{a}+b}\QATOPD[ ] {n+b}{i_{a}+b}=\QATOPD[ ] {n+b}{i_{a}+b}+(q^{n+b}-1)%
\QATOPD[ ] {n+b-1}{i_{a}+b-1}.
\end{equation}%
Therefore, once we write, 
\begin{equation}
q^{\frac{(i_{a}+b)(i_{a}+b-1)}{2}}\QATOPD[ ] {n+b}{i_{a}+b}=q^{-\frac{b(b+1)%
}{2}}q^{b(i_{a}+b)}\QATOPD[ ] {n+b}{i_{a}+b},
\end{equation}%
it is easy to see that 
\begin{equation}
\det \left( q^{\frac{(b-1)(2i_{a}+b-2)}{2}}\QATOPD[ ] {n+b-1}{i_{a}+b-1}%
\right) =q^{-\sum_{j}\frac{j(j-1)}{2}}\det \left( \QATOPD[ ] {n+b-1}{%
i_{a}+b-1}\right) .
\end{equation}%
Collecting everything we thus have

\begin{equation*}
\det \left( \QATOPD[ ] {n+b-1}{i_{a}+b-1}\right) =q^{\sum_{j}{\frac{j(j-1)}{2%
}}-\sum_{j}{\frac{i_{j}(i_{j}-1)}{2}}+n(\lambda )}\QATOPD[ ] {n}{\lambda }.
\end{equation*}%
From which we finally obtain: 
\begin{equation}
\boxed{ A_q(\lambda)=q^{n(\lambda')}. }
\end{equation}

\section{Notation for Chern-Simons quantities}

We give here some information about the Chern-Simons quantities that appear
in the text, mainly in $\left( \ref{20}\right) $. For more information, see 
\cite{Marino:2002fk} and references therein. To understand the origin of
other quantities in $\left( \ref{20}\right) $, one has to take into account
the constructions of Seifert homology spheres from surgery. Seifert homology
spheres can be constructed by performing surgery on a link $\mathcal{L}$ in $%
\mathbf{S}^{3}$ with $n+1$ components, consisting on $n$ parallel and
unlinked unknots together with a single unknot whose linking number with
each of the other $n$ unknots is one. The surgery data are $p_{j}/q_{j}$ for
the unlinked unknots, $j=1,\cdots ,n$, and 0 on the final component. $p_{j}$
is coprime to $q_{j}$ for all $j=1,\cdots ,n$, and the $p_{j}$'s are
pairwise coprime. After doing surgery, one obtains the Seifert space $M=X({%
\frac{p_{1}}{q_{1}}},\cdots ,{\frac{p_{n}}{q_{n}}})$. This is rational
homology sphere whose first homology group $H_{1}(M,\mathbf{Z})$ has order $%
|H|$, where 
\begin{equation}
H=P\sum_{j=1}^{n}{\frac{q_{j}}{p_{j}}},\,\,\,\,\,\mathrm{and}%
\,\,\,\,P=\prod_{j=1}^{n}p_{j}.  \label{orderh}
\end{equation}
Another topological invariant that will enter the computation is the
signature of $\mathcal{L}$, which turns out to be: 
\begin{equation}
\sigma (\mathcal{L})=\sum_{i=1}^{n}\mathrm{sign}\biggl( {\frac{q_{i}}{p_{i}}}%
\biggr) -\mathrm{sign}\biggl( {\frac{H}{P}}\biggr).  \label{signa}
\end{equation}
For $n=1,2$, Seifert homology spheres reduce to lens spaces, and one has
that $L(p,q)=X(q/p)$. For $n=3$, we obtain the Brieskorn homology spheres $%
\Sigma (p_{1},p_{2},p_{3})$ (in this case the manifold is independent of $%
q_{1},q_{2},q_{3}$). In particular, $\Sigma (2,3,5)$ is the Poincar\'{e}
homology sphere. Finally, the Seifert manifold $X({\frac{2}{-1}},{\frac{m}{%
(m+1)/2}},{\frac{t-m}{1}})$, with $m$ odd, can be obtained by integer
surgery on a $(2,m)$ torus knot with framing $t$. Note that in $\left( \ref%
{20}\right) $ the weight and root lattices of $G$ are denoted by $\Lambda _{%
\mathrm{w}}$ and $\Lambda _{\mathrm{r}}$, respectively.

Finally, there is a phase factor in $\left( \ref{20}\right) $ that comes
from the framing correction, that guarantees that the resulting invariant is
in the canonical framing for the three-manifold $M$. Its explicit expression
is: 
\begin{equation}
\phi =3\mathrm{sign}\left( \frac{H}{P}\right) +\sum_{i=1}^{n}12s\left(
q_{i},p_{i}\right) -\frac{q_{i}}{p_{i}},
\end{equation}%
where $\sigma (\mathcal{L})$ is again the signature of the linking matrix of 
$\mathcal{L}$ and $s(p,q)$ is the Dedekind sum: 
\begin{equation}
s(p,q)={\frac{1}{4q}}\sum_{n=1}^{q-1}\cot \Bigl({\frac{\pi n}{q}}\Bigr)\cot %
\Bigl({\frac{\pi np}{q}}\Bigr).
\end{equation}

\end{document}